\input harvmac
\input amssym

\tolerance=10000



%
\def\omit#1{}

\def\coeff#1#2{\relax{\textstyle {#1 \over #2}}\displaystyle}

\def\oneone{\rlap 1\mkern4mu{\rm l}}

\def\cL{{\cal L}} 
\def\cN{{\cal N}} 
 
\def\cR{{\cal R}}

\def\bfone{\relax{\rm 1\kern-.35em 1}}

%
\def\us{\bf}

\def\IC{\Bbb{C}}

\def\IP{\Bbb{P}}

\def\nup#1({Nucl.\ Phys.\ $\us {B#1}$\ (}
\def\plt#1({Phys.\ Lett.\ $\us  {#1B}$\ (}
\def\plt#1({Phys.\ Lett.\ $\us  {#1B}$\ (}
\def\cmp#1({Comm.\ Math.\ Phys.\ $\us  {#1}$\ (}
\def\prp#1({Phys.\ Rep.\ $\us  {#1}$\ (}
\def\prl#1({Phys.\ Rev.\ Lett.\ $\us  {#1}$\ (}
\def\prv#1({Phys.\ Rev.\ $\us  {#1}$\ (}
\def\mpl#1({Mod.\ Phys.\ Let.\ $\us  {A#1}$\ (}
\def\ijmp#1({Int.\ J.\ Mod.\ Phys.\ $\us{A#1}$\ (}
\def\atmp#1({Adv.\ Theor.\ Math.\ Phys.\ $\bf {#1}$\ (}
\def\cqg#1({Class.\ Quant.\ Grav.\ $\bf {#1}$\ (}
\def\jag#1({Jour.\ Alg.\ Geom.\ $\us {#1}$\ (}
\def\jhep#1({JHEP $\bf {#1}$\ (}

%



%
%
%
%
\lref\GauntlettSC{
J.~P.~Gauntlett, D.~Martelli, S.~Pakis and D.~Waldram,
``G-structures and wrapped NS5-branes,''
arXiv:hep-th/0205050.
}
%
\lref\GauntlettFZ{
J.~P.~Gauntlett and S.~Pakis,
``The geometry of D = 11 Killing spinors.''
JHEP {\bf 0304}, 039 (2003)
[arXiv:hep-th/0212008].
}
%
\lref\GauntlettWB{
J.~P.~Gauntlett, J.~B.~Gutowski and S.~Pakis,
``The geometry of D = 11 null Killing spinors,''
JHEP {\bf 0312}, 049 (2003)
[arXiv:hep-th/0311112].
}
%
\lref\MartelliKI{
D.~Martelli and J.~Sparks,
``G-structures, fluxes and calibrations in M-theory,''
Phys.\ Rev.\ D {\bf 68}, 085014 (2003)
[arXiv:hep-th/0306225].
}
\lref\GauntlettCY{
J.~P.~Gauntlett, D.~Martelli and D.~Waldram,
``Superstrings with intrinsic torsion,''
arXiv:hep-th/0302158.
}
%
\lref\GowdigereJF{
C.~N.~Gowdigere, D.~Nemeschansky and N.~P.~Warner,
``Supersymmetric solutions with fluxes from algebraic Killing spinors,''
arXiv:hep-th/0306097.
}
%
\lref\PilchUE{
K.~Pilch and N.~P.~Warner,
``N = 2 supersymmetric RG flows and the IIB dilaton,''
Nucl.\ Phys.\ B {\bf 594}, 209 (2001)
[arXiv:hep-th/0004063].
}
%
\lref\PilchJG{
K.~Pilch and N.~P.~Warner,
``Generalizing the N = 2 supersymmetric RG flow solution of IIB
supergravity,''
Nucl.\ Phys.\ B {\bf 675}, 99 (2003)
[arXiv:hep-th/0306098].
}
%
%
\lref\CorradoNV{
R.~Corrado, K.~Pilch and N.~P.~Warner,
``An N = 2 supersymmetric membrane flow,''
Nucl.\ Phys.\ B {\bf 629}, 74 (2002)
[arXiv:hep-th/0107220].
}
%
\lref\LeighEP{
R.~G.~Leigh and M.~J.~Strassler,
``Exactly marginal operators and duality in four-dimensional N=1
supersymmetric gauge theory,''
Nucl.\ Phys.\ B {\bf 447}, 95 (1995)
[arXiv:hep-th/9503121].
}
%
\lref\MaldacenaRE{
J.~M.~Maldacena,
``The large N limit of superconformal field theories and supergravity,''
Adv.\ Theor.\ Math.\ Phys.\  {\bf 2}, 231 (1998)
[Int.\ J.\ Theor.\ Phys.\  {\bf 38}, 1113 (1999)]
[arXiv:hep-th/9711200].
}
%
\lref\SeibergAX{
N.~Seiberg,
``Notes on theories with 16 supercharges,''
Nucl.\ Phys.\ Proc.\ Suppl.\  {\bf 67}, 158 (1998)
[arXiv:hep-th/9705117].
}
%
\lref\PopeJP{
C.~N.~Pope and N.~P.~Warner,
``A dielectric flow solution with maximal supersymmetry,''
arXiv:hep-th/0304132.
}
\lref\IBNW{
I.~Bena and N.~P.~Warner, {\it in preparation}.
}
%
\lref\KhavaevFB{
A.~Khavaev, K.~Pilch and N.~P.~Warner,
``New vacua of gauged N = 8 supergravity in five dimensions,''
Phys.\ Lett.\ B {\bf 487}, 14 (2000)
[arXiv:hep-th/9812035].
}
%
\lref\FreedmanGP{
D.~Z.~Freedman, S.~S.~Gubser, K.~Pilch and N.~P.~Warner,
``Renormalization group flows from holography supersymmetry and a
c-theorem,''
Adv.\ Theor.\ Math.\ Phys.\  {\bf 3}, 363 (1999)
[arXiv:hep-th/9904017].
}
%
%
\lref\PilchEJ{
K.~Pilch and N.~P.~Warner,
``A new supersymmetric compactification of chiral IIB supergravity,''
Phys.\ Lett.\ B {\bf 487}, 22 (2000)
[arXiv:hep-th/0002192].
}
\lref\PilchFU{
K.~Pilch and N.~P.~Warner,
``N = 1 supersymmetric renormalization group flows from IIB supergravity,''
Adv.\ Theor.\ Math.\ Phys.\  {\bf 4}, 627 (2002)
[arXiv:hep-th/0006066].
}
%
\lref\JohnsonZE{
C.~V.~Johnson, K.~J.~Lovis and D.~C.~Page,
``The Kaehler structure of supersymmetric holographic RG flows,''
JHEP {\bf 0110}, 014 (2001)
[arXiv:hep-th/0107261].
}
%
\lref\PopeBD{
C.~N.~Pope and N.~P.~Warner,
``An SU(4) Invariant Compactification Of D = 11 Supergravity On A Stretched Seven Sphere,''
Phys.\ Lett.\ B {\bf 150}, 352 (1985).
}
%
\lref\PopeJJ{
C.~N.~Pope and N.~P.~Warner,
``Two New Classes Of Compactifications Of D = 11 Supergravity,''
Class.\ Quant.\ Grav.\  {\bf 2}, L1 (1985).
}
%
\lref\KPNW{
K.~Pilch and N.~P.~Warner,
``$\cN=1$ Supersymmetric Solutions of IIB Supergravity from Killing Spinors,''
USC-04/02,   {\it to appear}.} 
%
%
\lref\WarnerDU{
N.~P.~Warner,
``Some Properties Of The Scalar Potential In Gauged Supergravity Theories,''
Nucl.\ Phys.\ B {\bf 231}, 250 (1984).
}
%
\lref\WarnerVZ{
N.~P.~Warner,
``Some New Extrema Of The Scalar Potential Of Gauged N=8 Supergravity,''
Phys.\ Lett.\ B {\bf 128}, 169 (1983).
}
%
\lref\NicolaiHS{
H.~Nicolai and N.~P.~Warner,
``The SU(3) X U(1) Invariant Breaking Of Gauged N=8 Supergravity,''
Nucl.\ Phys.\ B {\bf 259}, 412 (1985).
}
%
\lref\AhnAQ{
C.~h.~Ahn and J.~Paeng,
``Three-dimensional SCFTs, supersymmetric domain wall and renormalization
group flow,''
Nucl.\ Phys.\ B {\bf 595}, 119 (2001)
[arXiv:hep-th/0008065].
}
%
\lref\AhnBY{
C.~h.~Ahn and K.~s.~Woo,
``Domain wall and membrane flow from other gauged d = 4, n = 8  supergravity. I,''
Nucl.\ Phys.\ B {\bf 634}, 141 (2002)
[arXiv:hep-th/0109010].
}
%
\lref\AhnQG{
C.~h.~Ahn and K.~s.~Woo,
``Domain wall from gauged d = 4, N = 8 supergravity. II,''
JHEP {\bf 0311}, 014 (2003)
[arXiv:hep-th/0209128].
}
%
\Title{ \vbox{ \hbox{USC-04/01} 
\hbox{\tt hep-th/0403006} }} {\vbox{\vskip -1.0cm
\centerline{\hbox
{A Family of M-theory Flows }}
\vskip .5cm
\centerline{\hbox {with Four Supersymmetries }}
\vskip 8 pt
\centerline{
\hbox{}}}}
\vskip -0.9 cm
\centerline{Dennis Nemeschansky and  Nicholas P.\ Warner} 
\bigskip
\centerline{ {\it Department of Physics and Astronomy}}  
\centerline{{\it University of Southern California}} 
\centerline{{\it Los Angeles, CA
90089-0484, USA}}

\vskip 0.8cm
\centerline{{\bf Abstract}}
\medskip
We use the techniques of ``algebraic Killing spinors'' to obtain a family
of holographic flow solutions with four supersymmetries in M-theory. 
The family of supersymmetric backgrounds  constructed here includes the 
non-trivial flow to the $(2+1)$-dimensional
analog of the Leigh-Strassler fixed point as well as generalizations 
that involve the $M2$-branes spreading in a radially symmetric fashion
on the Coulomb branch of this non-trivial fixed point theory.  
In spreading out, these  $M2$-branes also appear to undergo dielectric
polarization into $M5$-branes. Our results
naturally extend the earlier applications of the ``algebraic Killing spinor''  method
and also  generalize  the harmonic Ansatz in that our entire family of new
supersymmetric backgrounds is characterized by the solutions of a single, 
second-order,  non-linear PDE.  We also show that our solution is a
natural hybrid of special holonomy and the ``dielectric deformation''
of the canonical supersymmetry projector on the $M2$ branes.
 
\vskip .1in
\Date{\sl {February, 2004}}

\parskip=4pt plus 15pt minus 1pt
\baselineskip=15pt plus 2pt minus 1pt

\newsec{Introduction}

The problem of finding, and classifying supersymmetric backgrounds with 
non-trivial $RR$-fluxes has been around for many years, but it is only relatively
recently that it has begun to be addressed systematically.    This problem has
taken on new significance because of the important role that fluxes play
in supersymmetry breaking backgrounds, and particularly in holographic RG flows.
The idea of $G$-structures has  provided a very useful classification framework 
and has led to new families of solutions
\refs{\GauntlettSC\GauntlettFZ\GauntlettCY\MartelliKI{--}\GauntlettWB}, 
but this approach is, as yet, not
computationally powerful enough to reproduce some of the 
physically important families of holographic RG flow solutions.    
There is, however, a closely related 
approach, that of algebraic Killing spinors,  which is somewhat more 
narrowly focussed, but appears to be computationally efficient
\refs{\GowdigereJF, \PilchJG}.   It is our purpose
here to further develop this technique by finding new families of $M$-theory 
flows with {\it four} supersymmetries.   Similar results may also be obtained 
in IIB supergravity \KPNW.

While the ideas presented here apply rather more generally, we will 
work within $M$-theory, and define the Killing spinors to be solutions of:
\eqn\gravvar{\delta \psi_\mu ~\equiv~ \nabla_\mu \, \epsilon ~+~ \coeff{1}{144}\,
\Big({\Gamma_\mu}^{\nu \rho \lambda \sigma} ~-~ 8\, \delta_\mu^\nu  \, 
\Gamma^{\rho \lambda \sigma} \Big)\, F_{\nu \rho \lambda \sigma} ~=~ 0 \,.}

The essential idea behind $G$-structures is to classify the special differential
forms that arise in supersymmetric flux compactifications.  With algebraic Killing
spinors, one tries to characterize the spin bundle of the supersymmetries directly
(and algebraically) in terms of the metric.    The relationship between these approaches
is very simple:  Given some Killing spinors, $\epsilon^{(i)}$, there are
associated differential forms:
\eqn\specialforms{\Omega^{(ij)}_{\mu_1 \mu_2 \dots \mu_k}  ~\equiv~ 
\bar \epsilon^{(i)} \, \Gamma_{\mu_1 \mu_2  \dots   \mu_k} \epsilon^{(j)} \,.}
Conversely, given all these forms one can reconstruct the spinors.  The whole 
point is that the differential forms satisfy systems of  first-order differential
equations as a consequence of \gravvar \GauntlettFZ.  In addition, Fierz identities
give detailed information about (partial) contractions of these 
differential forms, thereby relating them algebraically.  
Without background fluxes  one finds that these forms are 
harmonic, and that the two-forms often yield some K\"ahler or hyper-K\"ahler 
structure.   One is thus led rapidly the study of cohomology of complex manifolds.    
In the presence of fluxes the systems of differential equations, and the
algebraic relations between the forms is more complicated, but the resulting
$G$-structure is the natural generalization of the ideas of using cohomology and
complex structures in the absence of fluxes.

With algebraic Killing spinors the idea is solve \gravvar\ more directly.  In 
Calabi-Yau manifolds, or with $G_2$ structures one usually finds that the
Killing spinor bundles are trivially defined by an algebraic projection, involving
the special differential forms, applied to the complete spin bundle.   For the
simplest intersecting brane solutions these projections are the familiar 
projections parallel and perpendicular to the branes.  The idea is to generalize
this to make Ans\"atze for the projectors that define the Killing spinor bundles.
These projections determine the types of
differential forms that emerge from \specialforms, and so Ans\"atze for these 
projectors must be implicitly the same as selecting the type of
$G$-structure.   While it would be very interesting to pursue this line
of thought, our purpose here is more computationally oriented.  The issue
with algebraic Killing spinors is to determine the projector Ansatz, and to
do that it is valuable to construct explicit, and illustrative examples.
In \refs{\GowdigereJF, \PilchJG} this was done for families of flows with eight 
supersymmetries, while here we  are going to obtain and solve  natural Ans\"atze 
for projectors leading  to solutions  with four supersymmetries in M-theory.    
Among the family of solutions that we will
generate here is the $M$-theory flow \CorradoNV\  to a superconformal fixed point 
(with an $AdS_4$ background) with four supersymmetries \refs{\WarnerDU\WarnerVZ\NicolaiHS\AhnAQ\AhnBY{--}\AhnQG}.  We will see how 
this work naturally extends the ideas of \GowdigereJF, and elucidates the structure
further.  We will also see rather explicitly why previous classification schemes
have not led to these flows, and we also suggest how
to repair this omission.

As in \GowdigereJF, we will look for solutions that correspond to  distributions of branes 
(with additional fluxes) where the brane distribution depends
non-trivially, but arbitrarily  upon one ``radial'' variable, $v$.   The solutions therefore
depend upon two variables, $u$ and $v$, where $u$ is essentially a radial
coordinate transverse to the brane distribution.  Without the additional fluxes,
the solution would be elementary, and can easily be written in terms of a harmonic 
function, $H(u,v)$.  The presence of additional fluxes leads to a rather more complicated
family of solutions:  This family is also determined entirely by a solution, $g(u,v)$, of a 
second order PDE in $u$ and $v$, but this PDE is non-linear.  Asymptotically one has 
$g \to 0$ at large $(u,v)$, and the non-linear PDE has a very simple
perturbation expansion:
\eqn\gpert{g(u,v) ~=~ \sum_{n=1}^\infty \, g_n(u,v) \, \epsilon^n \,,}
where $\epsilon$ is a small parameter.  As in \GowdigereJF, one then show
 that $g_1$ satisfies a linear, homogeneous, second order PDE, while $g_n$ satisfies 
an equation with the same linear differential operator sourced by combinations of
$g_k$ for $k < n$.  In this way our solution generalizes the standard
harmonic Ansatz.

The first step is to identify projection operators that reduce  the dimension
of the relevant spinor space.  In \GowdigereJF\ this required two projectors to reduce
32   supersymmetries to eight, whereas here we have to construct three projection 
operators, $\Pi_j$, $ j=0,1,2$   to reduce the supersymmetries to four.  If the solution one
seeks is holographically dual to a theory with a Coulomb branch, then there will
be a non-trivial space of moduli for brane probes.  This moduli space will   be 
realized as either a conformally K\"ahler  (for four  supersymmetries) or conformally  
hyper-K\"ahler  (for eight  supersymmetries) section  of the metric.   On this
section of the metric the supersymmetries will satisfy projection conditions
$\Pi_j \epsilon =0$ where  the $\Pi_j$ have the elementary form:
\eqn\simprojone{\Pi_j ~=~ {1 \over 2}\,  \big( \oneone ~+~ \Gamma^{{\cal X}_j} \big)\,,}
where $\Gamma^{{\cal X}_j}$ denotes a product  of gamma-matrices parallel to the
moduli space of the branes.   For example, in \GowdigereJF\  there was one
such projector  and it implemented the
half-flat condition of the spinors on the hyper-K\"ahler moduli space.
For the flows considered here there is a six-dimensional K\"ahler moduli space and
to reduce to one-quarter supersymmetry one must isolate the spinor singlets
under the $SU(3)$ factor of the holonomy group.  As is familiar with Calabi-Yau 
$3$-folds, this can be achieved by requiring $\Pi_1 \epsilon= \Pi_2 \epsilon =0$ 
for two projectors of the form \simprojone, where ${{\cal X}_j}$ is chosen based
upon the complex structure.
Having found   $\Pi_1$ and $\Pi_2$, one conjectures that, in suitably chosen frames,
these projectors remain the same for the complete metric, and not just on
the conformally K\"ahler section.  The frames that accomplish this
extension to the complete metric are canonically determined because the $G$-structure 
associated with $\cN =1$ supersymmetry guarantees that the K\"ahler structure on 
the moduli space extends at least to an almost complex structure on the whole 
spatial part of the metric.

The subtlety is the last projector, $\Pi_0$, which, based
on the experience of \refs{\PopeJP, \GowdigereJF, \PilchJG}  is the deformation of the 
canonical projector:
\eqn\canonproj{\Pi_0 ~=~ {1 \over 2}\,  \big( \oneone ~+~ \cos \chi(u,v)\  \Gamma^{123} 
~+~ \sin \chi(u,v) \ \Gamma^{*} \big)\,,}
where $ \Gamma^{*} $ denotes a product (or sum of products) of  gamma-matrices
with $( \Gamma^{*} )^2 = \oneone$.
If  $\chi(u,v) \equiv 0$ then this the standard projector parallel to the
$M2$ branes.  One can also derive the form of  \canonproj\ by analyzing known
solutions \CorradoNV\ from gauged supergravity.  There is, however, a more
direct way to deduce the form of $\Gamma^{*}$: Based on 
\refs{\GowdigereJF, \PilchJG,\PopeJP,\IBNW},  we know that the form of \canonproj\ 
should be interpreted as a the result of the $M2$ branes becoming dielectrically
polarized into a distribution of $M5$ branes.  Thus $\Gamma^{*}$ must
be made up of terms of the form $ \Gamma^{123ABC}$ for some 
$A,B,C$.  This, combined with the fact that  $\Pi_0 $ must 
commute with $\Pi_1$ and $\Pi_2$, and with the symmetries, determines
the form of   \canonproj.

In section 2 we describe more precisely the class of holographic flows
that we are seeking, and then we make the complete Ansatz in section 3.
The Ansatz is then solved it section 4, where we show that 
the entire solution is generated by the solution to a single
PDE.   While the result is relatively simple, it is a rather complicated task in practice
to ``un-thread'' from \gravvar\ all the independent equations for the Ansatz functions.
However, there are some short-cuts that can be made
using some of the simplest of the $G$-structure equations.  This is described
in section 5, where we focus the role of the almost complex
structure.  The surprise is that this  structure
only exists in ten dimensions: It requires one to include the two spatial
coordinates of the brane. We believe that this fact goes some way to explain why 
previous analyses  have not led to the families of solution
described here:  The deformations in  \canonproj\ lead to almost complex structures
on the entire spatial part of the metric, and not merely on the ``internal''
space.

We conclude by drawing out general ideas from the results presented here
and in earlier work.  We also describe some general open problems to which
our methods might find useful application.

\newsec{The holographic flows}

The holographic theory on $M2$-branes is an  $\cN=8$ supersymmetric theory with 
eight scalars,  eight fermions and sixteen supercharges.   The $AdS_4 \times S^7$ 
background yields the holographic dual of a strongly coupled, superconformal
fixed point  \refs{\MaldacenaRE,\SeibergAX}.  One can also think of this field theory as 
arising through a  limit of the Kaluza-Klein reduction of $\cN=4$ supersymmetric 
Yang-Mills theory on a circle.    The extra scalars in three-dimensions come from the
components of the gauge fields along the circle and a Wilson line
parameter around the circle.   From this perspective one can frequently link
results obtained for Yang-Mills theory to results for the three-dimensional
scalar-fermion theory.

A particular example of such a link is the flow to a new superconformal fixed point, with
four supersymmetries, obtained by giving a mass to a single chiral multiplet.
For $\cN=4$ Yang-Mills theory this flow was analyzed by Leigh and Strassler \LeighEP\
and its holographic dual was identified in
\refs{\KhavaevFB\FreedmanGP\PilchEJ{--}\PilchEJ}.  Being strongly coupled, it
is very hard to work with the scalar-fermion theory directly, however there is
an exactly corresponding holographic flow  that was described in \CorradoNV.  
To be specific, the eight scalars and fermions can be paired into four complex
superfields, $\Phi_k$, $k=1,\dots,4$.  Giving one of them, say, $\Phi_4$, a mass
leads to an $\cN=2$ supersymmetric  flow  (four supersymmetries) to a new 
superconformal fixed point.  The vevs of remaining scalars, $\Phi_k$, $k=1,\dots,3$,
parametrize  the Coulomb branch at this fixed point.  A brane-probe analysis
\JohnsonZE\ of the supergravity solution does indeed reveal a three complex-dimensional
space of moduli for the brane probes, and that this moduli space has, as one
should expect, a natural K\"ahler structure.

It is this family of flows that we seek to generalize here. The flow of \CorradoNV\ was obtained
from gauged supergravity and leads to {\it a single point} on the Coulomb branch
of the fixed point theory.  From the field theory and the brane-probe analysis
 we know that there should be a family of solutions parametrized by a function
 of six variables describing a general brane distribution on the Coulomb branch
 at the fixed point.  As a first step to finding this general class, we are going to 
 seek flow solutions that correspond to brane distributions with rotational symmetry
 on the Coulomb branch:  That is, we will seek solutions where the brane
 density depends only upon a radial coordinate on the moduli space.

\newsec{The Ansatz}

\subsec{Conventions}

Our $M$-theory conventions are those of \refs{\PopeBD, \PopeJJ}.  The metric is 
``mostly plus,'' and we take the gamma-matrices to be
\eqn\gammamats{\eqalign{& \Gamma_1 ~=~  -i \, \Sigma_2 \otimes 
\gamma_9 \,, \quad
\Gamma_2~=~  \Sigma_1 \otimes \gamma_9 \,, \quad 
\Gamma_3~=~   \Sigma_3 \otimes \gamma_9 \,, \cr &
\Gamma_{j+3} ~=~  {\oneone}_{2 \times 2} \otimes \gamma_j \,, 
\quad j=1,\dots,8 \,,}}
where the $\Sigma_a$ are the Pauli spin matrices, $\oneone$ is
the Identity matrix, and  the $\gamma_j$ are real, symmetric $SO(8)$
gamma matrices.  As a result, the $\Gamma_j$ are all real, with 
$\Gamma_1$ skew-symmetric and $\Gamma_j$ symmetric for $j>2$.  One also
has:
\eqn\prodgamma{
\Gamma^{1 \cdots\cdots 11} ~=~ \oneone \,,}
where $\oneone$ will henceforth denote the $32 \times 32$ identity matrix.
The gravitino variation will be as in \gravvar.   With these conventions,
sign choices and normalizations,  the equations of motion are:
\eqn\eqnmot{\eqalign{ R_{\mu \nu} ~+~ R \, g_{\mu \nu}  ~=~&  \coeff{1}{3}\, 
F_{\mu \rho \lambda \sigma}\, F_\nu{}^{ \rho \lambda \sigma}\,,\cr 
\nabla_\mu F^{\mu \nu \rho  \sigma} ~=~& -  \coeff{1}{ 576} \, \varepsilon^{\nu \rho  \sigma
\lambda_1\lambda_2
\lambda_3 \lambda_4 \tau_1\tau_2  \tau_3 \tau_4} \,F_{ \lambda_1  \lambda_2
 \lambda_3  \lambda_4} \, F_{ \tau_1\tau_2 \tau_3 \tau_4} \,.}}

\subsec{Background fields}

We take the metric to have the general form;
\eqn\metansatz{\eqalign{ds^2_{11}  ~=~ &  e^{2\, A_0} \,  ( -dx_0^2 +   dx_1^2 + 
dx_2^2) ~+~ e^{2\, A_1} \, (du^2 ~+~ u^2\, d\phi^2)~+~ e^{2\, A_2} \, dv^2   \cr & ~+~ 
 v^2 \, e^{2\, A_3} \, \big(d \lambda^2 ~+~ \coeff{1}{4}\, \sin^2 \lambda (\sigma_1^2  +
\sigma_2^2)  ~+~\coeff{1}{4}\, \sin^2 \lambda \,  \cos^2 \lambda \,\sigma_3^2 \big)
\cr &  ~+~v^2\, e^{2\, A_2}\big(\, e^{  A_4} \big(d\psi -\coeff{1}{2} \,\sin^2 \lambda \, 
\sigma_3 \big)~+~ \,e^{ A_5} \, d\phi \,\big)^2   \,,}}
where $A_0, \dots , A_5$ are, as yet, arbitrary functions of $u$ and $v$.
This Ansatz is a natural generalization of the metric in \CorradoNV, and 
the $(u,v)$  coordinates are related to the variables of \CorradoNV\ via:
\eqn\uvvars{u~=~ e^{A(r)} \, \sqrt{\sinh\chi(r)} \, sin \theta \,, \qquad
v~=~ e^{{1 \over 2} A(r)} \, \rho(r)  \, sin \theta \,.}
We describe how we arrived at this change of variable at the end of section 5.
The important aspect of this change of variables is that it
results in conformally flat metric in the $(u,\phi)$ directions, and as we will
describe below, it reveals that the complete spatial section of the
metric has an almost complex structure. 

One should also note that $(u,\phi)$ 
parametrize the internal directions {\it transverse} to the brane
moduli space, and that  $v$, $\psi$, $\lambda$  and the $\sigma_j$ sweep
out the moduli space.    The middle line of \metansatz\ is the metric
on $\IC\IP^2$, with an $SU(3)$ isometry, and as was noted in \CorradoNV,
this part of the metric could be replaced by any Einstein-K\"ahler manifold.
The radial coordinate on the brane moduli space is thus $v$, and the solutions we 
seek have brane distributions that depend solely upon $v$, and by symmetry,  the 
metric and gauge fields must  depend only upon $u$ and $v$. 

We use the frames: 
\eqn\framansatz{\eqalign{e^j  ~=~ &  e^{A_0} \, dx_{j-1} \,, \ \ j=1,2,3\,, \qquad
e^4  ~=~    e^{A_1} du  \,,  \qquad  e^5  ~=~    e^{A_2} dv  \,,   \qquad
e^6  ~=~    v\, e^{A_3} \, d \lambda  \,, \cr   
e^{6+j}   ~=~  &  \coeff{1}{2}\,  v\, e^{A_3}  \,  \sin  \lambda\,   \sigma_{j} \,, \ \ j=1,2 \,, \qquad
e^{9}   ~=~   \coeff{1}{2}\,  v\, e^{A_3}  \,  \sin  \lambda  \,  \cos  \lambda \, \sigma_{3} \,,  
\cr 
e^{10}  ~=~  &  v \, e^{ A_2}\big(\, e^{  A_4} \big(d\psi -\coeff{1}{2} \,\sin^2 \lambda \, 
\sigma_3 \big)~+~ \,e^{ A_5} \, d\phi \,\big) \,, \qquad   e^{11}  ~=~    u\, e^{A_1}\,  d\phi
  \,.}}
In \CorradoNV\  it was found that the tensor gauge field had a natural holomorphic
structure on the internal space if it is written in terms of the frames.  We therefore
respect this structure in making the Ansatz:
\eqn\formansatz{\eqalign{ A^{(3)} ~=~  p_0(u,v) \, e^1 \wedge e^2 \wedge e^3  
 & ~+~  {\rm Re} \big[ e^{i(\phi+3\,\psi)} \big(p_1(u,v) \, ( e^4 + i \, e^{11}) \cr & ~+~ 
p_2(u,v) \, ( e^5 + i \, e^{10}) \big) 
\wedge  ( e^6 -i \, e^{9}) \wedge ( e^7 -i \, e^{8}) \big] \,. }}
The holomorphic pairing of frames is straightforward.  First recall
that the positions in the internal space can be interpreted in terms
of vevs of the holomorphic scalars, $\Phi_k$.  In the flows we
seek the branes are spreading in the $\Phi_1, \Phi_2,\Phi_3$ directions,
while $\Phi_4$ is the field that is given a mass.    The latter corresponds
to the $(u,\phi)$ directions, and has a manifest complex structure in 
\metansatz.  In the remaining directions there is $\IC\IP^2$ with a K\"ahler
structure that  leads to the $( e^6 -i \, e^{9}) \wedge ( e^7 -i \, e^{8}) $ term
in   \formansatz, while the remaining frames form the last pair.
In section 5 we  will  see how the  this holomorphic structure is guaranteed
by the $G$-structure, but for the present we note that the K\"ahler
structure on the brane-probe moduli space must be a conformal
multiple of:
\eqn\modKstr{J_{moduli} ~\equiv~ e^6 \wedge e^9 ~+~ e^7 \wedge e^8 ~-~
e^5 \wedge e^{10} \,.}

\subsec{Projectors}

Having identified the moduli space and found the  complex structure on it, 
one is immediately led to the following projectors:
\eqn\simprojs{\Pi_1 ~=~ {1 \over 2}\,  \big( \oneone ~+~ \Gamma^{6789} \big) \,,
\qquad \Pi_2 ~=~ {1 \over 2}\,  \big( \oneone ~-~ \Gamma^{ 578\,10} \big)  \,.}
The third projector, $\Pi_3~=~ {1 \over 2}\,  \big( \oneone ~-~ \Gamma^{ 569\,10} \big)$,
is redundant given the other two.
The choices of signs in these projectors are set by the choices of signs in the 
complex factors of \formansatz.  Note that the $(4,11)$ index-pair is absent.   
These two projectors thus act in directions parallel to the brane moduli space.
Imagine restricting the spinors to this slice of the metric:  The brane moduli
space has a K\"ahler $3$-fold, and so it natural to reduce the spinors in terms
of the $SU(3)  \times U(1)$ holonomy, and isolate the supersymmetry
 as being the $SU(3)$ singlet.    Imposing the requirement that 
$\Pi_1  \epsilon = \Pi_2 \epsilon  = 0$ implements this.  One can 
also see that this is equivalent to imposing the $SO(6)$ helicity conditions:
\eqn\helconds{\Gamma^{69} \, \epsilon ~=~  \Gamma^{78}\, \epsilon ~=~  
- \Gamma^{5\,10} \, \epsilon ~=~  \pm i\, \epsilon \,.}
The form of these projectors on the moduli space is thus required by
the underlying K\"ahler structure.  The leap that we make in the general
Ansatz is to assume, as we did in \GowdigereJF\  that the projectors are unmodified
as one moves off the space of moduli, and are given globally through the
almost complex structure by \simprojs.

 The non-trivial task is to find the deformation of the standard $M2$-brane
 projector.  We take this to be:
\eqn\hardproj{\eqalign{\Pi_0 ~=~ & {1 \over 2}\,  \Big( \oneone ~+~ 
{2\,\kappa \over 1 + \kappa^2}\, 
 \Gamma^{123}  ~+~  {1- \kappa^2  \over 1 + \kappa^2}\, \big( \sin(\phi + 3 \psi)  
 \Gamma^{4567\,11} +  \cos(\phi + 3 \psi)  \Gamma^{4568\,11}  \big) \Big) \cr
 ~=~& {1 \over 2}\,  \Big( \oneone ~+~  {2\,\kappa \over 1 + \kappa^2}\, 
 \Gamma^{123}  ~-~  {1- \kappa^2  \over 1 + \kappa^2}\, \Gamma^{123} \, \big( 
 \sin(\phi + 3 \psi)   \Gamma^{89 \,10} +  \cos(\phi + 3 \psi)  \Gamma^{79\,10}  
 \big) \Big) \,.}}
where $\kappa=\kappa(u,v)$ is an arbitrary function\foot{We have  
chosen to use rational parametrization of the circular functions $\cos\chi$
and $\sin \chi$ in \canonproj.  This is largely because {\it Mathematica$^{TM}$}
is more efficient with rational expressions.}, and
the second identity follows from $\Gamma^{1 \dots 11} =\oneone$,

While this projector seems rather complicated, it is in fact relatively simple to understand.
One can argue simplify from the mathematical structure, but there is 
a very useful piece of physical input:  The interpretation of \hardproj\ \refs{\PopeJP,\IBNW} 
should be associated with dielectric polarization of the $M2$ branes into $M5$ branes.
This means that the deformation of the projector should
involve the $M5$-brane component $\Gamma^{123ABC}$ for some choice of $A,B,C$.
Next, observe that $\Pi_0$ must commute with $\Pi_1$ and $\Pi_2$, and this
is achieved in \hardproj\   by having $A,B,C$ be  exactly one out of each of the 
``complex pairs'': $(6,9)$, $(7,8)$, $(5,10)$.  There are $2^3 = 8$ such choices, however,  
the projection conditions,  $\Pi_1 \epsilon = \Pi_2 \epsilon  = 0$, imply that only two of 
those choices are really  independent, for example $\Pi_1 \epsilon  = 0$ means that 
 $\Gamma^{67} \epsilon = \Gamma^{89} \epsilon$ {\it etc.}.    The two independent
 choices are those matrices appearing in \hardproj.  
 
 Finally, to understand the
 angular dependence in \hardproj\ one needs to use the fact that the projectors
 must commute with the global symmetry generators.
 
 \subsec{Isometries and Lie derivatives}
 
 The metric has an $SU(3) \times U(1)^2$ isometry acting on the internal
 space.  The $SU(3)$ acts transitively on the $\IC\IP^2$ directions, while  the
 $U(1)$'s are translations in $\phi$ and $\psi$.  The tensor gauge field
 is obviously invariant under  under a combination of these $U(1)$'s but it is,
 in fact, invariant under   the complete $SU(3) \times U(1) \times U(1)$ group.  
 To see the $U(1)$ invariances  more explicitly, use the following expressions for 
 the $\sigma_j$:
\eqn\oneforms{\eqalign{\sigma_1 ~\equiv~&  \cos \varphi_3\, d\varphi_1 ~+~ 
\sin\varphi_3\, \sin\varphi_1\, d \varphi_2 \,, \cr
\sigma_2 ~\equiv ~&  \sin\varphi_3\, d\varphi_1 ~-~ 
\cos\varphi_3\, \sin\varphi_1\, d \varphi_2 \,, \cr
\sigma_3 ~\equiv ~&  \cos\varphi_1\, d\varphi_2 ~+~   d \varphi_3  \,,}}
 from which one finds:
 \eqn\holbit{e^7 - i\,e^8 ~\sim~ e^{-i \varphi_3} \,( d\varphi_1 ~+~ i\, 
 \sin\varphi_1\, d \varphi_2) \,.}
 Thus a translation in $\psi$ or $\phi$ can be compensated in \formansatz\ by a
 translation in $\varphi_3$, or a rotation in the $(7,8)$-plane of the frames.
 This means that the $U(1)$ invariances of the background are generated by
 the Killing vectors, $L_{(1)}^\mu$ and $L_{(2)}^\mu$, where:
 \eqn\Kvecs{L_{(1)}^\mu \, \partial_\mu ~\equiv~ {\partial \over \partial \varphi_3}  
 ~+~  {1\over 3} \, {\partial \over \partial \psi}  \,, \qquad 
 L_{(2)}^\mu \, \partial_\mu ~\equiv~ {\partial \over \partial \phi}  
 ~-~   {1\over 3}  \, {\partial \over \partial \psi}    \,.}

Given a Killing isometry of the metric, one can define the Lie derivative of a spinor field:
 \eqn\LieDeriv{\eqalign{{\cL}_K \, \epsilon ~\equiv~ & K^\mu\, \nabla_\mu \, \epsilon
 ~+~ \coeff{1}{4} \,  (\nabla_{[\mu} \, K_{\nu ]})\, \Gamma^{\mu \nu} \, \epsilon \cr ~=~ &
 K^\mu\, \partial_\mu \, \epsilon  ~+~ \coeff{1}{4} \,  (K^\rho \, \omega_{\rho \mu \nu} ~+~ 
 \nabla_{[\mu} \, K_{\nu ]})\, \Gamma^{\mu \nu} \, \epsilon\,.}}
The factors are fixed by the requirement that this reduce to the usual
Lie derivative on the vector $\bar \epsilon \Gamma^\mu \epsilon$.

A  geometric symmetry of the solution must either act trivially on
the Killing spinors, or it must be an $\cR$-symmetry.  For our flow, there
is only a single $U(1)$ $\cR$-symmetry, and this can determined which  
through perturbative analysis.  We find that $ L_{(2)}$ generates the non-trivial
 $\cR$-symmetry while the Killiing spinor must be a singlet under the
 action of $L_{(1)}$.   It is also important to note that the Killing spinor
 must be a singlet under the $SU(3)$ isometry.  
 This means that
 \eqn\Kactions{{\cL}_K \, \epsilon ~\equiv~ 0 \,.}
when  $K$  is either $L_{(1)}$ or any $SU(3)$ generator.  

It is trivial to compute the terms that make up the Lie derivative from
the metric \metansatz, and one finds that almost all the terms
cancel between the connection and the curl of $K_\mu$ in \LieDeriv.  
Indeed, one finds that:
$$
 {\partial \over \partial y^\mu}  \, \epsilon ~=~ 0\,, \qquad y^\mu \in \{\lambda, \varphi_1, 
 \varphi_2, \varphi_3 \}  \,. 
 $$
 That is, the Killing spinor must be independent of the $\IC\IP^2$ directions. From 
 the $L_{(1)}$ action one obtains:
 \eqn\Loneact{{\partial \over \partial \psi}  \, \epsilon ~=~ - \,{3 \over 2} \, \Gamma^{78} 
 \, \epsilon\,.}
 Moreover. if $\epsilon^{(a)}$, $a =1,2$ are the two Killing spinors, the 
 $L_{(2)}$ action yields:
 \eqn\Ltwoact{\eqalign{\Big( {\partial \over \partial \phi} ~-~ {1 \over 3}\, {\partial \over \partial \psi}
 \Big)  \, \epsilon^{(1)} &  ~=~   - \, \epsilon^{(2)} \cr 
 \Big( {\partial \over \partial \phi} ~-~ {1 \over 3}\, {\partial \over \partial \psi}
 \Big)  \, \epsilon^{(2)} &  ~=~   + \, \epsilon^{(1)}\,.}}

Having properly identified the symmetry actions on the spinors, one can then fix
the complete angular dependence of the projection operator, $\Pi_0$,
by requiring that it commute with these Lie derivative operators.
 
\newsec{The new solutions}

One now simply inserts the Ansatz into the gravitino variation, \gravvar\ and tries to solve
all the equations.    The system is hugely overdetermined, but at first sight is a little
overwhelming.    There are, however, some very useful simplifications.

First, there are combinations of the gravitino variations in which the tensor
gauge fields cancel, leaving only metric terms:
\eqn\simpcombs{ \Gamma^1 \delta \psi_1 ~+~ \Gamma^7 \delta \psi_7
~+~ \Gamma^8 \delta \psi_8 ~=~ 0\,, \qquad 
\Gamma^1 \delta \psi_1 ~+~ \Gamma^6 \delta \psi_6
~+~ \Gamma^9 \delta \psi_9 ~=~ 0\,.}
From these one immediately learns that
\eqn\firstmess{\eqalign{\partial_u\,(A_0 + 2\, A_3) ~=~ & 0\,, \qquad
\partial_u\,(A_4 + 2\, A_2  -  2\, A_3 ) ~=~ 0\,,  \cr 
v\, \partial_v\,(A_0 + 2\, A_3) ~=~ & 2\, ( 1 ~-~ e^{2\,A_2 - 2\,A_3 + A_4})   \,.}}
From this we see that $(A_4 + 2\, A_2  -  2\, A_3 )$ is a function
of $v$ alone, however, because $e^5 =  e^{ A_2} \,dv$, there is the freedom 
to re-define $A_2$ up to an abritrary function of $v$, and hence arrange that 
$(A_4 + 2\, A_2  -  2\, A_3 ) = 0$.  It then follows that $(A_0 + 2\, A_3) = const.$.
We can scale  the $x_j$ and thereby set this constant to zero, but we will
preserve it as an explicit scale.  Thus these simple combinations
of variations lead to:
\eqn\firsteqns{  A_0   ~=~  - 2\, A_3 ~+~ 2 \log(2\, L) \,, \qquad
 A_4  ~=~ -2\, (A_2  -  A_3)  \,,} 
where $L$ is a constant scale factor.

After this the solution proceeds a little more slowly.  The most direct procedure
is to observe that there are only eight independent functions of $u$ and $v$
in the frame components, $F_{abcd}$, of the field strength.  One writes 
the variations in terms of these eight functions, and then carefully cross 
eliminates.  This leads to the equations:
\eqn\pzeroeqn{  p_0   ~=~  - {\kappa \over 1+ \kappa^2}    \,,  \qquad
{1- \kappa^2  \over 1+ \kappa^2}   ~=~  c \, u\, e^{4 \, A_3 - A_1}  \,,} 
where $c$ is a constant of integration.  Using these facts, and continuing
the cross-elimination yields:
\eqn\ptwoeqn{  p_2   ~=~  - {1- \kappa \over 1+ \kappa}    \,, }
\eqn\secondeqn{ v\, \partial_v\,(A_1 -  A_2) ~=~   3\, (  e^{2\,(A_2 -  A_3) }  ~-~  1) \,.} 
At this point, the equations become rather complicated, but it
is possible to disentangle them to obtain the following:
\eqn\poneeqn{ p_1   ~=~  - \, {2\, v \over u}\,  {(1- \kappa^2) \over (1+ \kappa^2)} 
\    e^{- A_1 + A_2 + A_5 } \,,} 
\eqn\thirdeqn{ u\, \partial_u\,(A_1 -  A_2) ~=~   - 3\,   e^{2\,A_2 -  2\, A_3 + A_5}   \,,} 
\eqn\fourtheqn{\partial_u\,(A_1 -  A_2) ~=~   \, {3 \, v \over 4\, u}\,  \partial_v
\,\bigg[{(1+  \kappa)^2 \over  (1+  \kappa^2)}  \  e^{ 2\, (A_1 -    A_3)} \bigg]  \,,} 
\eqn\fiftheqn{u\,  \partial_u \,\bigg[{(1+  \kappa)^2 \over  (1+  \kappa^2)}
\  e^{ 2\, (A_1 -    A_3)} \bigg]  ~=~   - 2\,   {(1-  \kappa)^2 \over  (1+  \kappa^2)}
\  e^{ 2\, (A_1 -    A_3)}   \,.} 

This system of equations suffices to solve everything.  Define:
\eqn\ghdefn{  g \equiv 2\, (A_1 -  A_2) \,, \qquad  h \equiv   \,  
{(1+  \kappa)^2 \over  (1+  \kappa^2)}  \  e^{ 2\, (A_1 - A_3)}  \,,}
then the foregoing imply:
\eqn\partone{\partial_u\, g ~=~ {3 \over 2} \, {u \over v} \, \partial_v \, h \,, }
\eqn\parttwo{\eqalign{u\, \partial_u\, h ~-~ 2\, h  ~=~ & -4\,  e^{ 2\, (A_1 -    A_3)}
\cr  ~=~ & - {2 \over 3} \, {1 \over v^5} \, \partial_v \big( v^6\, e^g \big) \,.}}
Eliminating $h$ from these two equations yields the ``master equation'' for $g$:
\eqn\master{u^3\, \partial_u\, \big(  u^{-3} \,   \partial_u \, g \big)~+~  v^{-1}
\, \partial_v\, \big(  v^{-5}  \, \partial_v \big( v^6\, e^g \big) \big) ~=~ 0 \,.}
This may also be written:
\eqn\newmaster{ \partial^2_u\,   g  ~-~  3\, u^{-1} \partial^2_u\,   g ~+~  
\big(\, \partial^2_v\,   g   ~+~ 7\, v^{-1}\, 
 \partial_v\, g ~+~  ( \partial_v\, g)^2 \,\big) \, e^g   ~=~ 0 \,.}

Suppose that one has a solution, $g$, to this equation; that is, one knows 
$(A_1 -  A_2)$.  One can then obtain $(A_2-   A_3)$ and $ A_5$ from \secondeqn\ and 
\thirdeqn\ by differentiating $g$.  From \fourtheqn\ and \parttwo\ one can then
obtain $\partial_v (u^{-2} h)$ and $\partial_u (u^{-2} h)$ respectively, and by 
quadrature one  thus obtains $\kappa$.  From this and \pzeroeqn\ one obtains $4 A_3 - A_1$,
and hence $A_1, A_2$ and $A_3$ independently.  Finally, the $p_j$ and
$A_0$ and $A_4$ are obtained from \pzeroeqn, \poneeqn, \ptwoeqn\ and
\firsteqns.  In this procedure almost every function is obtained directly
from $g$ and its derivatives.  The only exception is that $h$ is obtained
by quadrature.  Thus having solved \master, all other parts of the
solution are obtained by elementary operations:  At no point do we have 
to solve any further equations.  It is in this sense that \master\ is the
master equation of the new family of solutions.
 
So far we have solved the Killing spinor equations.  As was pointed in 
\refs{\GauntlettFZ,\GowdigereJF} this is not enough to guarantee that the 
solution we found also satisfies the equation  of motion.   The reason for this 
is that the commutator of two supersymmetries does not necessarily generate 
{\it all} the equations of motion.   An analysis of precisely
which subset of the equations of motion are generated may
be found in \GauntlettFZ, and using this one can significantly reduce the
number of equations that need to be verified. 

On the other hand, the hard work has largely been done in solving
the supersymmetry variations, and the difficulty of verifying all the 
equations of motion is not overly burdensome, and so we checked them
all for completeness.    With  \metansatz\  there are eight non-trivial Einstein equations. 
There are six independent diagonal terms\foot{Here we are using the
Poincar\'e symmetry on the brane and the $SU(3)$ symmetry on the
$\IC\IP^2$.} (with    $ \mu= \nu $), and two non-diagonal ones with 
$ (\mu=4 , \nu=5)$ and  $ (\mu=10 ,  \nu=11)$. 
All of these Einstein equations can be expressed in terms of the relations 
\firsteqns--\master\ found by solving the Killing  spinor equations or their $u$ and $v$ 
derivatives, and so our solution satisfies all the Einstein equations.

The Maxwell equations can be verified in  a similar fashion. There are many more 
terms to be checked but again after lengthy but straight forward algebraic 
manipulations  one finds, once again,  that the Maxwell  equations can be expressed 
in terms of the relations  \firsteqns--\master\  and their $u$ and $v$ derivatives.
Thus  the PDE \master\ is both sufficient and necessary for solving all the equation of motion.

\newsec{$G$-structures and simplifications}

The foregoing analysis was done by working directly with the
supersymmetry transformations.  In practice it is non-trivial 
to extract the simple equations from \gravvar.  This entire process 
can be sometimes be simplified by use of some of the $G$-structure identities.

First, the vectors
\eqn\Kvecs{K_{(ij)}^{\mu }  ~\equiv~  \bar \epsilon^{(i)} \, \Gamma^{\mu} \epsilon^{(j)} \,,}
are always Killing vectors, and usually one finds that if these are non-zero
then they are translations to the brane.  There are examples where one
generates other Killing vectors \refs{\GowdigereJF, \PilchJG, \IBNW}, but here 
we find that the only  Killing vectors obtained from \Kvecs\ are indeed those along 
the brane.

The other very useful identities involve the $2$-forms:
\eqn\specids{  \Omega^{(ij)}_{\mu \nu}      ~=~   \epsilon^{(i)} \, \Gamma_{\mu \nu}
 \epsilon^{(j)}  \,,}
which satisfy   simple differential identities as a  consequence of \gravvar.  If one skew 
symmetrizes such an identity, it reduces to:
\eqn\extderOm{  \partial_{[\rho} \Omega^{(ij)}_{\mu \nu ]}     ~=~   F_{\rho \mu \nu \sigma}
\, K_{(ij)}^\sigma   \,.}
An immediate application of this to observe that if the Killing vector components,
$K_{(ij)}^{\mu } $, are in fact constant, and the field $ F_{\rho \mu \nu \sigma}$
is independent of the coordinates dual to these Killing vectors, then there is
a gauge in which:
\eqn\specgauge{  \Omega^{(ij)}_{\mu \nu}     ~=~ \coeff{1}{3}\,  A^{(3)}_{\mu \nu \rho }
\, K_{(ij)}^\rho   \,.}
This determines some of the components of $A^{(3)}$, and in particular, leads 
immediately from the projector \hardproj\  to the expression for $p_0$ in 
\pzeroeqn.

Suppose now that $\epsilon^{(i)} = \epsilon^{(j)} = \epsilon$, and define
\eqn\spectens{K^{\mu }  ~\equiv~  \bar \epsilon  \, \Gamma^{\mu} \epsilon \,, 
\qquad \Omega _{\mu \nu}  ~\equiv~  \bar \epsilon  \, \Gamma_{\mu\nu} \epsilon  \,.}
One can then derive \GauntlettFZ\ the algebraic identities:
\eqn\specids{ {\Omega_\mu} ^{\rho} \, {\Omega_\rho}^{\nu}   ~=~   (K^2) \, \delta^\nu_\mu
~-~ K_\mu \, K^\nu \,, \qquad K^\mu  \Omega _{\mu \nu}  ~=~   0\,.}
Indeed, by suitable choice of $\epsilon$, one can arrange that $K^\mu$ is
simply $K^\mu = \delta^\mu_0$, and hence:
\eqn\spatial{ {\Omega_b} ^{c} \, {\Omega_c}^{a}   ~=~  e^{2  A_0} \, \delta^a_b \,,
\qquad {a, b, c = 2, \dots, 11} \,.}
Therefore, the ten-dimensional, spatial metric is has an almost complex structure, 
and the obvious question is whether it is a complex, or even K\"ahler structure.   
More precisely, if one conformally rescales the metric by $e^{-2  A_0} $ then
$\Omega _{\mu \nu}$ provides an almost complex structure satisfying 
\extderOm, which means that it is closed except for components parallel
to the brane.

If one uses the metric Ansatz \metansatz\ and imposes the projection conditions using
\simprojs\ and \hardproj, then the $2$-form, $\Omega$, becomes:
\eqn\twoform{\eqalign{ \Omega  ~=~ &  e^{A_0} \,\Big[\,   {2\, \kappa \over 1 + \kappa^2} \,  
(e^2 \wedge e^3 ~+~ e^4 \wedge e^{11}) ~+~ e^6 \wedge e^9 ~+~  e^7 \wedge e^8   ~-~ 
 e^5 \wedge e^{10} \cr 
  ~+~& {1-\kappa^2 \over 1 + \kappa^2} \, ( (- \cos(2\,\phi) \, e^2 +  \sin(2\,\phi) \, e^3)
 \wedge e^4 + ( \cos(2\,\phi) \, e^3 +  \sin(2\,\phi) \, e^2)
 \wedge e^{11}) \,\Big] \,   \,.}}
Observe that this contains and extends the K\"ahler structure of the moduli space
 \modKstr\ to the entire spatial section of the metric.  The fact that the $G$-structure
 must contain such a form, $\Omega$,  gave us the canonical way to extend the
 special holonomy projectors away from the moduli space.
 One should also note how the non-trivial deformation of the projector $\Pi_0$ 
manifests itself here in terms of a nontrivial mixing, or fibering of the spatial directions 
of the brane over the $(u,\phi)$ directions.  Moreover, this mixing is precisely what 
spoils the  global  K\"ahler structure:  One cannot project $\Omega$ perpendicular
to the branes and preserve the projection of \spatial. 

One can also derive many useful identities from \twoform\ and 
\extderOm.  As we have already remarked, the equations parallel to the brane
lead to $p_0$ as given in \pzeroeqn.  The components parallel to the $\IC\IP^2$
factor lead a system like \firstmess, and which reduces to \firsteqns\ after fixing
the remaining coordinate invariance.    The components in the directions
$(u,\phi,x^j)$ and $(v,\phi,x^j)$ then lead to the second equation in \pzeroeqn.
Finally, the closure in the $(u,v, \phi)$ direction leads to the differential equation:
\eqn\mixedDE{u\,  \partial_v \,\bigg[{2\,\kappa  \over  (1+  \kappa^2)}
\  e^{ 2\, (A_1 -    A_3)} \bigg]  ~+~  v\,  \partial_u \,
\big(  e^{ 2\,  A_2 -   2\, A_3 + A_5 } \big) ~=~0   \,,} 
which is a combination of the equations found in section 4.

The construction of the forms, $\Omega^{(ij)}$, was also very useful in finding
the coordinate transformations \uvvars\ for the solution of \GowdigereJF.  In particular,
by constructing \twoform\ and projecting parallel to the brane one can read
off the complex pair of differentials $(du,d\phi)$.

\newsec{Final Comments}

We have shown that the ``algebraic Killing spinor'' technique can
be adapted to problems with four supersymmetries.   In this paper, and
in \GowdigereJF\ we were able to deduce the projectors that define
the Killing spinors through a combination of two ideas.  First, one looks
at the supersymmetry conditions on the brane-probe moduli
space, where the metric is either K\"ahler, or hyper-K\"ahler.   On this
space the supersymmetries are obtained via standard holonomy
techniques, and the resulting projectors can be extended to the complete 
(spatial) metric via the almost complex structure.     The second idea is that while 
the original holographic theory is based on some standard brane configuration,
the non-trivial deformation involves a dielectric deformation of this
configuration into some five-branes.  The canonical supersymmetry projector
must reflect this deformation, and its form can be fixed from this and 
the fact that it must commute with the symmetry generators and
with the other projectors.

We have now obtained quite a number of families of solutions that
are motivated by physically important RG flows and yet involve 
this dielectric deformation of the underlying branes \refs{\GowdigereJF,
\PilchJG, \KPNW, \PopeJP,\IBNW}.   These solutions all involve a moduli space 
of branes, typically a Coulomb branch of some non-trivial flow, and they describe 
the spreading of the branes in a symmetric manner.  These solutions thus generalize  
the usual harmonic brane Ansatz.    Indeed, one is led to solutions that are
characterized by a single function, but that function is
determined by a non-linear PDE.   Here the ``master equation'' is
\master, while in \GowdigereJF\ it was:
\eqn\basicPDE{  {1 \over u^3}\, {\del \over \del u} \Big( \,u^3 {\del \over \del u}
\Big( {1 \over u^2} \,  f(u,v)  \Big) \Big) ~+~  {1 \over v}  \,
{\del  \over \del v} \Big(  f(u,v)  \, {1 \over  v}\,  {\del \over \del v}\,
\Big( {v^2  \over u^2}  \,  f(u,v)  \Big) \Big) ~=~ 0\,.}
At first sight this seems somewhat different, but one can reduce to an
equation of the same form as \master\ by setting $f(u,v) =e^{g(u,v)}$ and
then canceling a factor of $e^{g(u,v)}$ from each side of the equation.
While these equations are non-linear, they do have a relatively simple, linear
perturbation expansion of the form \gpert\ \GowdigereJF.  In particular, at
leading order there is a ``seed function'' that satisfies a homogenous, linear
PDE that presumably encodes the distribution of branes on the moduli-space.

More generally, our projector  Ansatz \simprojs\ and \hardproj\ represents
a rather simple modification of the corresponding Ansatz on a Calabi-Yau
manifold, particularly given the appearance of a ten-dimensional almost complex 
structure in our solution.  

The evolution of of the work presented here, and in \refs{\GowdigereJF,
\PilchJG,  \KPNW,  \PopeJP,\IBNW},  is perhaps somewhat reminiscent of the story of
``warp factors.''   The latter originally emerged from the technicalities of understanding 
how  lower-dimensional supergravity theories were embedded within 
higher-dimensional theories, but in recent years, warp factors have entered into
the mainstream of string compactification, and even into phenomenology.   In
this paper, and our earlier ones, we started by trying to understand how 
supersymmetry was broken in a class of solutions coming from lower-dimensional
supergravity only to find that, in higher dimensions, these solutions can be generalized
to whole families and that these families all necessarily involve  dielectric 
deformations of the original branes.  It would therefore not surprise us if
the classes of solutions we are considering will be of rather broader
significance in the study of supersymmetry breaking and 
supersymmetric backgrounds.

\bigskip
\leftline{\bf Acknowledgements}
 
This work was supported in part by funds
provided by the DOE under grant number DE-FG03-84ER-40168.
N.W. would like to thank Iosif Bena for helpful discussions.

\listrefs
\vfill
\eject
\end